\documentclass[journal=jacsat,manuscript=article]{achemso}

\usepackage{graphicx}
\usepackage{placeins}
\usepackage{amssymb}
\usepackage{subcaption}
\usepackage[font=small,labelfont=bf]{caption}
\usepackage{tabularx} 
\usepackage{float}
\usepackage{csquotes}
\usepackage[version=3]{mhchem} 

\author{Milad Entezami}
\email{milad.entezami@uwaterloo.ca}
\affiliation[1]
{Department of Electrical and Computer Engineering, University of Waterloo, Waterloo, Ontario N2L 3G1, Canada}
 \author{Seyed Ali Hosseini Farahabadi}
 \affiliation[1]
{Department of Electrical and Computer Engineering, University of Waterloo, Waterloo, Ontario N2L 3G1, Canada}
 \author{Man Chun Alan Tam}
 \affiliation[1]
{Department of Electrical and Computer Engineering, University of Waterloo, Waterloo, Ontario N2L 3G1, Canada}
\author{Andree Coschizza}
\affiliation[2]{Department of Physics and Astronomy, University of Waterloo, Waterloo, Ontario N2L 3G1, Canada}
\author{Jan Kycia}
\affiliation[2]{Department of Physics and Astronomy, University of Waterloo, Waterloo, Ontario N2L 3G1, Canada}
\author{Zbigniew R. Wasilewski}
\email{zbig.wasilewski@uwaterloo.ca}
\affiliation[1]{Department of Electrical and Computer Engineering, University of Waterloo, Waterloo, Ontario N2L 3G1, Canada}
\alsoaffiliation[2]{Department of Physics and Astronomy, University of Waterloo, Waterloo, Ontario N2L 3G1, Canada}
\alsoaffiliation[3]{Waterloo Institute for Nanotechnology (WIN), University of Waterloo, Waterloo, Ontario N2L 3G1, Canada}
\alsoaffiliation[4]{Institute for Quantum Computing (IQC), University of Waterloo, Waterloo, Ontario N2L 3G1, Canada}

\title
     {Strain-Balanced Low-Temperature-Grown Beryllium-Doped InGaAs/InAlAs Superlattices for High-Performance Terahertz Photoconductors under 1550~nm Laser Excitation}

\abbreviations{}
\keywords{strain-balanced Superlattices (SLs), Low-Temperature Be-Doped InGaAs/InAlAs, 1550~nm-Based Photoconductive Terahertz Antenna, Ultrafast Carrier Dynamics}

\begin{document}








\begin{abstract}
This study systematically investigates the photoconductive properties of low-temperature-grown Beryllium (Be)-doped InGaAs/InAlAs strain-balanced superlattices (SLs) grown by molecular beam epitaxy under stationary growth conditions on semi-insulating InP:Fe substrates. The stationary growth approach enabled precise control over lateral gradients in layer strain, composition, and thickness across a single wafer, while strain-balancing facilitated pseudomorphic growth to explore a wide range of structural parameters, providing a robust platform to study their influence on photoconductive performance. Structural characterization confirmed high crystalline quality and smooth surface morphology in all samples.
Time-resolved pump–probe spectroscopy revealed subpicosecond carrier lifetimes, validating the effectiveness of strain balancing and Be doping in tuning ultrafast recombination dynamics. Hall effect measurements supported by 8-band k.p modeling revealed enhanced carrier mobility in strain-balanced SLs compared to lattice-matched structures, primarily due to reduced electron and hole effective masses and stronger quantum confinement. Additionally, optical absorption under 1550~nm excitation showed improved absorption coefficients for the strain-balanced structure, consistent with the reduction in bandgap energy predicted by theoretical modeling, thereby enhancing photon-to-carrier conversion efficiency. Furthermore, transmission electron microscopy provided first-time evidence of significant Be-induced interdiffusion at the strained SL interfaces, an important factor influencing carrier transport and dynamics. These findings position low-temperature-grown Be-doped InGaAs/InAlAs strain-balanced SLs as promising materials for high-performance broadband THz photoconductive detectors operating at telecom-compatible wavelengths. 

\end{abstract}

\section{Introduction}
During the past decade, terahertz (THz) science and technology have gained significant interest due to the unique non-ionizing properties and superior penetration capabilities of THz radiation, which enable it to interact with optically opaque materials~\cite{Int1}. These features make THz applications invaluable in diverse fields such as wireless communication~\cite{Int2, Int3}, security screening~\cite{Int4}, biomedical imaging~\cite{Int5,Int6}, and non-destructive material testing, particularly for thickness measurements and defect identification via imaging and spectroscopy methods~\cite{Int7,Int8}. 

Photoconductive antennas (PCAs) lie at the core of terahertz time-domain spectroscopy (THz-TDS) systems, one of the most commonly used state-of-the-art pulsed THz generation and detection methods. Illuminating a photoconductive material with femtosecond optical laser pulses generates ultrafast photoexcited carriers, which accelerate through biased PCAs, leading to the generation of a transient photocurrent. This photocurrent can then be coupled to the metallic nanostructures of the PCAs, resulting in the radiation of THz waves~\cite{int9}. In recent years, THz-TDS has transitioned from a purely scientific tool to an emerging technology with industrial applications. Despite considerable advancements, research continues to focus on developing high-power, efficient THz sources and broadband THz detectors with high dynamic ranges. Recent efforts have aimed to enhance THz-TDS systems by replacing bulky and expensive Ti-Sapphire femtosecond lasers with compact, reliable, and cost-effective fiber lasers operating at a telecom-compatible wavelength of 1550~nm. Although fiber-laser-driven THz-TDS systems are already functional, further performance improvements remain highly desirable.

The ternary compound InGaAs has been shown to be the most suitable photoconductive material for fiber-laser-driven THz generation, allowing the development of portable pulsed THz spectroscopy and imaging systems~\cite{Int10}. However, bulk InGaAs suffers from high intrinsic carrier concentrations, leading to low dark resistivity. It also exhibits relatively long photoexcited carrier lifetimes, two distinct factors that limit its performance. To meet the criteria for high-performance photoconductive THz antennas, the materials must exhibit~\cite{Int11}:
\begin{itemize}
    \item A high absorption coefficient at 1550~nm to ensure efficient carrier generation.
    \item High dark resistivity to suppress dark current under applied DC bias. 
    \item High carrier mobility to enhance the dynamic range of the THz signal. 
    \item Subpicosecond photoexcited carrier lifetime for broad THz bandwidths.  
\end{itemize}

Balancing these properties is inherently challenging, as strategies that shorten carrier lifetime, such as introducing point defects, also reduce mobility by increasing carrier scattering. Precise material engineering is therefore required, and molecular beam epitaxy (MBE) offers a powerful platform for fine-tuning these competing parameters toward optimized THz sources and detectors compatible with fiber laser excitation.

An effective strategy involves engineering superlattice (SL) structures by alternating layers of the higher bandgap InAlAs barriers with InGaAs quantum wells. The growth of these SLs at temperatures around 400~°C promotes a degree of phase separation in the InAlAs barriers due to the interplay of surface kinetics and thermodynamics. This leads to the formation of InAs-rich and AlAs-rich domains within the InAlAs alloy~\cite{oh1990dependence}. The localized trap states introduced by these defects within the bandgap increase the dark resistivity to $ \rho=2500~\Omega\text{.cm}$ while maintaining high electron mobility $\mu_e= 2700 ~ \text{cm}^2\text{/Vs}$ of InGaAs layers~\cite{dietz2011thz}. However, the carrier lifetime remains relatively long, $ \tau > 40~\text{ps}$, thus limiting bandwidth performance~\cite{Int12}.
At low temperature (LT) MBE growth ($<$ 300~°C) of InGaAs / InGaAs SLs, excess arsenic incorporation into the InGaAs lattice leads to the formation of arsenic antisite defects ($\text{As}_{\text{Ga}}$), which act as electron traps and significantly influence the photoconductive properties~\cite{globisch2016terahertz}. Adding beryllium (Be) doping ionizes these defects and compensates the n-type conductivity. This ionization enhances dark resistivity and reduces carrier lifetimes by rapidly capturing electrons into ionized defect states, but it comes at the cost of decreased carrier mobility~\cite{Int22}. By optimizing the doping concentration of Be, subpicosecond carrier lifetime $ \tau <1 ~ \text{ps}$ was demonstrated, making them suitable for THz photoconductive detectors. However, shortening the lifetime came at the cost of a lower resistivity of $ \rho=339 ~\Omega\text{.cm}$, and a lower electron mobility of $\mu_e= 335 ~ \text{cm}^2\text{/Vs}$  ~\cite{Int13, Int21}. 

Another method involves doping the InGaAs crystal lattice with transition metals. MBE growth at 400~°C with optimized iron (Fe) doping improves resistivity to $ \rho= 532 ~\Omega\text{.cm}$ with a higher electron mobility of $ \mu_e = 900~ \text{cm}^2\text{/Vs}$~\cite{Int14}. Rhodium (Rh) doping has been reported to give even better performance, producing resistivity of $ \rho=3190~\Omega\text{.cm}$  and electron mobility of $ \mu_e=1010 ~ \text{cm}^2\text{/Vs}$~\cite{Int15}. However, maintaining the necessary high doping concentration requires an effusion cell temperature of approximately 1800~°C, which presents significant practical challenges~\cite{Int16}. Alternatively, embedding ErAs precipitates within InGaAs/InAlAs SLs introduces deep-level states that pin the Fermi level near mid-gap. This approach achieved a resistivity of $ \rho=600 ~\Omega\text{.cm}$ and electron mobility of $ \mu_e= 1100 ~ \text{cm}^2\text{/Vs}$~\cite{Int11}, while recent work demonstrates enhanced band engineering by incorporating aluminum (Al) and bismuth (Bi) into an InGaAlBiAs matrix, further optimizing dark resistivity and carrier lifetimes at 1550~nm excitation \cite{acuna2024band}. Recent studies also showed that undoped InGaAs/InAlAs SLs grown at 400~°C on GaAs substrates with metamorphic buffers show significantly reduced carrier lifetimes under 800~nm femtosecond laser excitation, likely due to the very high density of threading dislocations~\cite{Int18}. However, a detailed characterization of their photoconductive properties is still lacking.

The examples above are only part of the growing volume of work in this area, which illustrates the importance of portable and cost-effective THz TDS technology.

In this study, we propose extending the search for the optimum materials for THz TDS to include the so-called strain-balanced In$_x$Ga$_{1-x}$As/In$_y$Al$_{1-y}$As SLs, where the In content $x$ in the wells is considerably higher than that needed for the lattice matching to the InP substrate. The resulting compressive stress is perfectly compensated by tailored tensile stress introduced into the barrier, bringing the net stress to zero within each period of the SL.  This allows the growth of an arbitrary number of SL repeats without exceeding the critical thickness, thus avoiding the generation of a detrimental high density of misfit dislocations. We discuss how structural parameters of such SLs, including strains, quantum well and barrier compositions, and layer thicknesses, impact photoconductive properties. We present design considerations as well as low-temperature MBE growth and characterization of Be-doped InGaAs/InAlAs strain-balanced SLs on InP substrates. Leveraging the unique capabilities of MBE, we grew these SLs on stationary (i.e., non-rotating) substrates. This method takes advantage of substantial yet well-characterized variations of individual In, Ga, and Al atomic fluxes across the stationary wafer, allowing for the exploration of broad lateral variations in the SL's structural parameters. A comprehensive investigation involving time-resolved pump-probe spectroscopy, transport Hall effect measurements, optical absorption, and nanoscale material characterization was undertaken to aid in finding the optimal structural parameters for enhancing THz detector performance at 1550~nm femtosecond laser pulse illumination.

The structure of this paper is organized as follows: \textbf{Section I, \enquote{Strain Engineering and Electronic State Modeling}} outlines the theoretical framework for calculating the structural parameters essential to achieving strain-balanced InGaAs/InAlAs SL growth without lattice relaxation. \textbf{Section II, \enquote{Stationary Low-Temperature MBE Growth}} describes the growth process, highlighting its methodology and advantages.  \textbf{Section III: \enquote{Material Characterization}} presents the techniques used to evaluate crystal quality, including high resolution X-ray diffractometry (HRXRD), scanning transmission electron microscopy (S/TEM), Nomarski differential interference contrast (DIC) microscopy and atomic force microscopy (AFM). Finally, \textbf{Section IV: \enquote{Photoconductive Properties}} evaluates the photoconductive responses of the grown SLs through time-resolved pump-probe spectroscopy, transport Hall effect measurements, and optical absorption spectroscopy.

\section{Strain Engineering}

In heteroepitaxial growth, when a sufficiently thin epitaxial layer is grown on a much thicker substrate with a slight lattice mismatch,  interatomic forces will apply biaxial stress to the layer, forcing its in-plane lattice constant ($a_{\parallel}$) to align with the substrate lattice constant, creating lateral strain $\epsilon_{\parallel} $ in the layer.  Such in-plane strain is defined as: 
\begin{equation}
 \epsilon_{\parallel}=\frac{a_{sub}-a_{layer}}{a_{layer}}
  \label{eqn:strain}
\end{equation}
To minimize elastic energy, the vertical lattice constant ($a_{\perp}$) expands for compressive strain  ($\epsilon_{\parallel} < 0$) or contracts for tensile ($\epsilon_{\parallel} > 0$) strain,  resulting in a tetragonal distortion of the unit cell.  As a consequence, the electronic properties of such a strained layer change, opening the possibility of their deliberate tailoring through what is called strain engineering. However, maintaining high crystal quality under strained conditions poses challenges. As strain increases, the maximum layer thickness for pseudomorphic epitaxy decreases, beyond which layer quality can rapidly deteriorate as a result of misfit dislocation generation.  This critical thickness at which dislocations start to form to relieve excess stress can be estimated using the Matthews–Blakeslee equation, providing a conservative limit on how thick any individual strained layer can be before inelastic deformation occurs~\cite{matthews1975defects}.

When multiple strained layers are stacked during heteroepitaxial growth, each below its critical thickness, stress can accumulate, ultimately triggering the formation of misfit dislocations and degrading the performance of the device. 

\subsection{Strain-Balanced Superlattices}

A powerful technique to mitigate stress accumulation is \textit{strain-balancing}, in which alternating layers under compressive and tensile strains effectively compensate the net stress, preserving pseudomorphic epitaxy and preventing relaxation. This technique requires a precise selection of the compositions and thicknesses of the epitaxial layers in each SL period.  

For considered here  In$_x$Ga$_{1-x}$As/In$_y$Al$_{1-y}$As SLs, nominal compositions $x$ and $y$ reported in the literature were chosen as 0.53 and 0.52, respectively. Such a choice ensures lattice matching of every layer to the InP substrate, allowing strain-free epitaxial growth of SLs thick enough to have substantial 1550-nm absorption of the femtosecond laser pump photons.  

In the present study, we propose to increase the light absorption by decreasing the band gap of InGaAs in the SL's quantum wells. This requires an increase in the indium mole fraction $x$ of the wells, which puts them under compressive stress ($x>0.53$). To ensure pseudomorphic growth with an arbitrary number of SL repeats, we meet the strain-balancing condition within each SL period by growing In$_y$Al$_{1-y}$As barriers under tensile stress ($y<0.52$), ensuring that net in-plane stress ${\bar\sigma_{\parallel}}$ is negligible across each SL period satisfies the condition: 
\begin{equation}
    \bar{\sigma_{\parallel}}=\frac{(\sigma_w\,l_w+\sigma_b\,l_b)}{(l_w+l_b)}\approx0.
\end{equation}
where $\sigma_{w}$ and $\sigma_{b}$ represent the in-plane strains of the InGaAs wells and the InAlAs barriers, respectively, while $l_{w}$ and $l_{b}$ denote their corresponding thicknesses. 

\begin{figure} [t!]
 \centering
 \includegraphics[width=0.6\textwidth] {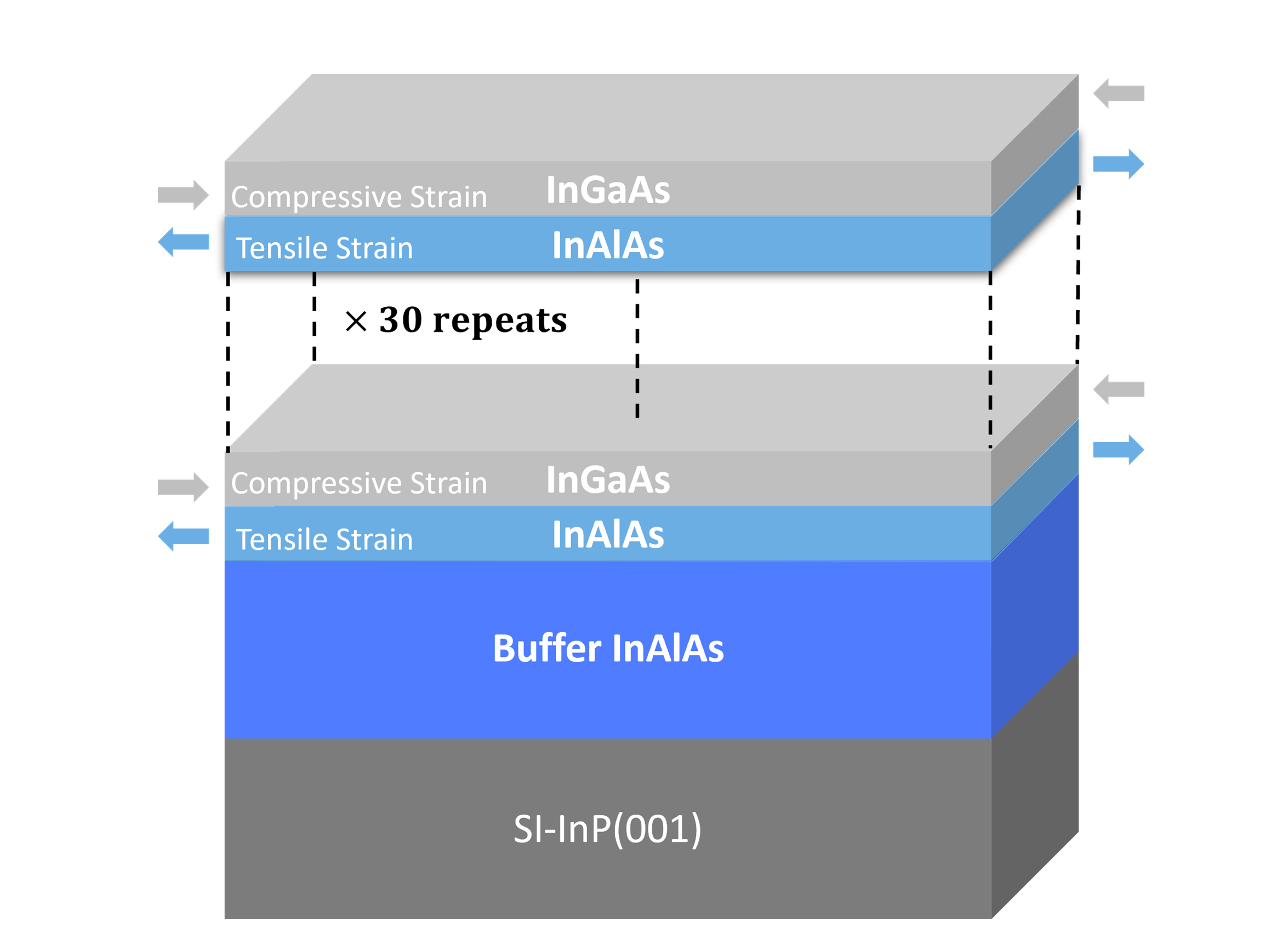}
 \caption{Schematic diagram of strain-balanced SLs with compressively strained wells and tensile-strained barriers.}
  \label{fig:0}
\end{figure}

A schematic representation of this structure is shown in Figure \ref{fig:0}.  
Theoretically, the strain-balancing condition is governed by the relationship between the lattice constants of the substrate, the quantum wells, and the barriers, along with their corresponding thicknesses, by minimizing the average elastic energy in the structure~\cite{harrison2016quantum}:

\begin{equation}
  a_{InP} = \frac{\sum_{k=1}^{n} A_k l_k / a_k}{\sum_{k=1}^{n} A_k l_k / a_k^2} =\frac{ A_w l_w / a_w + A_b l_b / a_b}{ A_w l_w / a_w^2 + A_b l_b / a_b^2}
  \label{eqn:strain-balanced}
\end{equation}
Here, $a_{w}$ and $a_{b}$ represent the lattice constants of the InGaAs wells and InAlAs barriers, respectively, while $l_{w}$ and $l_{b}$ denote their corresponding thicknesses. The elastic stiffness constant for each layer, $A_{w}$ and $A_{b}$, are defined as: 

\begin{equation}
  A_{w(b)} = C_{11}+C_{12}-2\frac{C_{12}^2 }{C_{11}}
  \label{eqn:stiffness}
\end{equation}
where $C_{i,j}$ represents the stiffness coefficient for InGaAs and InAlAs along the [001] crystal direction. To obtain the needed material parameters, we use Vegard's law,  i.e. linearly interpolate lattice parameters and stiffness constants between the GaAs, AlAs and InAs end compounds.~\cite{vurgaftman2001band}   With this, the lattice constant for In$_x$Ga$_\text{1-x}$As is given by $(5.65338 + 0.405x)$~\AA, while that for In$_y$Al$_\text{1-y}$As is $(5.66143 + 0.391y)$~\AA. The elastic stiffness constants $C_{11}$, $C_{12}$, and $C_{44}$  for In$_x$Ga$_\text{1-x}$As, are expressed as $(122.1 - 38.81x)$~GPa, $(56.6 - 11.34x)$~GPa, and $(60 - 20.41x)$~GPa, respectively. Similarly, for In$_y$Al$_\text{1-y}$As, the respective values are given by $(125 - 41.71y)$~GPa, $(53.4 - 8.14y)$~GPa, and $(54.2 - 14.61y)$~GPa. 

\begin{figure}
\centering
\includegraphics[width=0.70\textwidth]{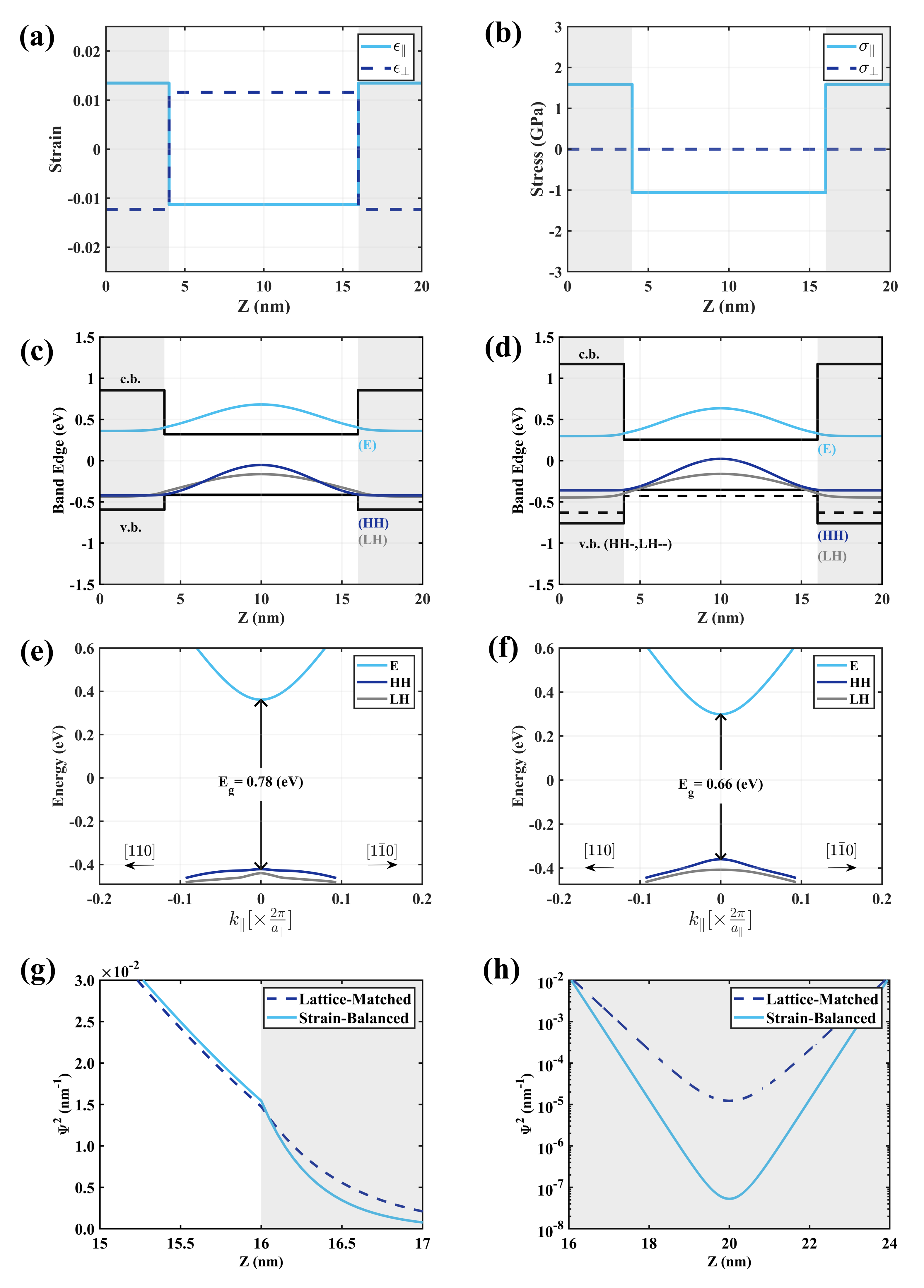}
\caption{(a) calculated in-plane and out-of-plane strain ($\epsilon_{\parallel},\epsilon_{\perp} $) and (b) in-plane and out-of-plane stress ($\sigma_{\parallel},\sigma_{\perp}$) distributions in infinitely periodic strain-balanced SLs with $x=0.7$ and $y=0.325$, shown for a single quantum well. (c) and (d) also present the calculated band-edge profiles and ground-state wave functions (spatial probability densities \(\Psi^2(z)\)) corresponding to electrons in the first conduction subband (E) and hybridized heavy-hole (HH) and light-hole (LH) valence subbands for lattice-matched and strain-balanced SLs, respectively. (e) and (f) show the energy dispersion relations (E(k)) in two orthogonal in-plane crystal directions [$110$] and [$1\bar{1}0$] calculated using the 8-band k.p method for lattice-matched and strain-balanced SLs, respectively. (g) and (h) compare the electron wave-function penetration profiles at the quantum well-barrier interface and in the barrier layer for both strain-balanced and lattice-matched single quantum wells, shown on linear and logarithmic scales, respectively. The gray-shaded regions indicate the barriers layers.}
\label{fig:1}
\end{figure}

Once the thicknesses of the wells and the barriers, together with the target mole fraction $x$ of indium in the In$_x$Ga$_{1-x}$As wells, are defined, the corresponding strain-balancing composition $y$ of the In$_x$Al$_{1-x}$As barriers can be determined using equation (\ref{eqn:strain-balanced}). Figure~\ref{fig:1}.a-b present an example of in-plane and out-of-plane strain profiles and in-plane stress profile for a single strain-balanced quantum well structure (12~nm wells, 8~nm barriers) representing one SL period.  We also plot the out-of-plane stress profile to emphasize that it is zero. Indeed, no stress is applied in this direction.

\subsection{Electronic State Modeling}

This section discusses the electronic states modeling for periodic strain-balanced and lattice-matched In$_x$Ga$_{1-x}$As/In$_y$Al$_{1-y}$As SLs grown on SI-InP(001):Fe substrates. Numerical simulations were performed using the nextnano Schrödinger-Poisson quantum solver in the 8-band k.p approximation. These simulations were used to assess the impact of epitaxial strain on the electronic states of the superlattices. Periodic boundary conditions were applied to both sides of a single quantum well to approximate infinite SL periods. The well and barrier thicknesses were set to 12~nm and 8~nm, respectively.

The band edge profiles and ground state wavefunctions for electrons (E), as well as hybridized heavy-hole (HH) and light-hole (LH) valence subbands in the lattice-matched and strain-balanced SLs, are shown in Figure~\ref{fig:1}.c and~\ref{fig:1}.d, respectively. The conduction band (c.b.) and valence band (v.b.) profiles reveal notable differences between the two cases. The energy dispersion relations are illustrated in Figure~\ref{fig:1}.e and~\ref{fig:1}.d, for lattice-matched and strain-balanced SLs, respectively. In strain-balanced SLs, the conduction and valence band edges are shifted, resulting in a reduced bandgap energy of 0.66 (eV), compared to 0.78 (eV) in lattice-matched SLs. This bandgap reduction is advantageous for enhancing optical absorption at 1550~nm.  Another important effect is the increased curvature of the upper valence band at k$_{\parallel}$=0, which translates into a significantly lower in-plane effective mass of the holes, thus their higher mobility. Furthermore, the spatial probability densities, $\Psi^2(z)$, of the electron, hybridized heavy-hole, and light-hole wave functions indicate stronger quantum confinement in the strain-balanced SLs, thus reducing scattering by InAlAs alloy fluctuations.  Indeed, as shown in Figure~\ref{fig:1}.g-h, the penetration depth of the ground state electron wave function into the barriers is reduced in strain-balanced SLs due to a significantly higher barrier for these quantum wells. However, these higher barriers and a slight increase in $\Psi^2(z)$ at the interfaces will likely lead to an increase in interfacial scattering, possibly counteracting the mobility-enhancing effect of the above mechanisms. 

For the calculations presented here, we do not include the effects of Be doping or the final thickness of the superlattices.  These factors are difficult to evaluate for real structures. Beryllium is a shallow donor, and in principle, evaluating the hole concentration profiles and related modifications to the band edges and wave functions would be straightforward.  However, since the growths were carried out at low temperature, there are unknown high levels of antisite As$_\text{Ga}$ deep donors present in the structures.  Moreover, their exact concentrations are expected to be different in the barriers and wells, considerably altering the carrier distribution profiles.  Other important factors that can significantly alter the band edge profiles in finite structures are the Fermi level pinning at the surface and at the interface with the InP substrate. These will be discussed later. 

\section{Stationary Be-doped Low-Temperature MBE Growth}

All growths reported here were performed on quarters of 3" SI-InP(001):Fe substrates using a Veeco GEN10 MBE system. The heterostructures consist of a 700~nm InAlAs buffer layer nominally lattice matched to an InP substrate and 30 periods of Be-doped compressively strained InGaAs and tensile strained InAlAs layers, forming strain-balanced SLs with a total thickness of 600~nm. To explore a wide range of structural parameters within a single growth, we used a stationary growth technique~\cite{wasilewski1991studies,wasilewski1997composition}, stopping the rotation of the wafer during the deposition of superlattice layers. 

The design of this MBE system and effusion cells ensures the uniformity of the thickness of the layers better than 2\% on 3" substrates, when grown with substrate rotation.  However,  the individual atomic fluxes incident on a wafer decrease by a factor of approximately 1.8 across a 3" substrate, from the closest point to the furthest point away from a given effusion cell.  The spatial distributions of the fluxes are reproducible and can be accurately characterized with dedicated calibration structures. Thus, stopping the wafer rotation for the growth of a layer will transfer this flux distribution to a well-controlled variation of the layer thickness.  

\begin{figure}[t!]
\centering
\includegraphics[width=\textwidth]{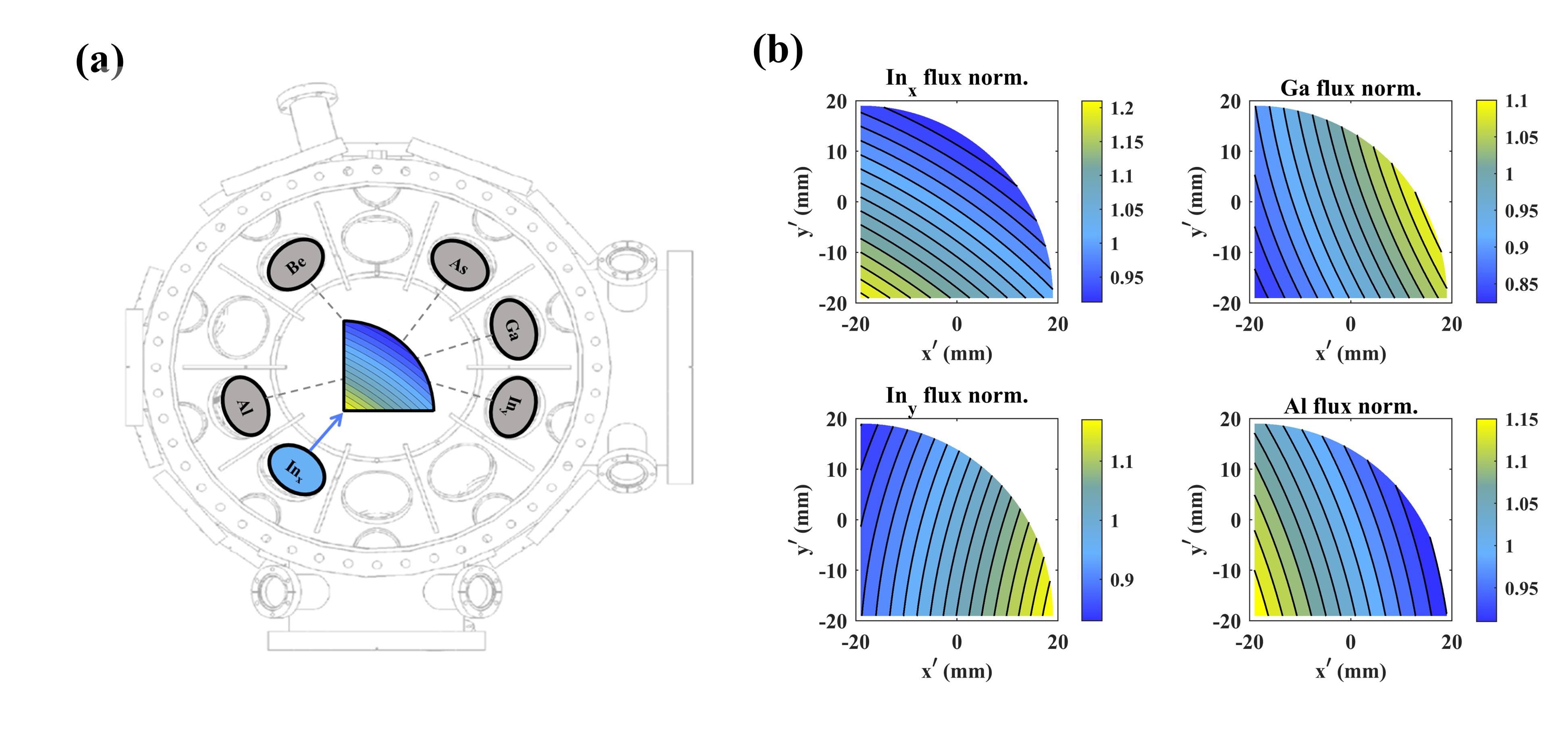}
\caption{(a) Schematic representation of the MBE growth chamber configuration and material fluxes from effusion cells at different azimuthal positions. (b) Normalized incoming flux distribution on the sample.}
\label{fig:2}
\end{figure}

The In, Al and Ga flux distributions over the stationary substrates used here are shown in Figure~\ref{fig:2}.b. Because of the different azimuthal locations of the effusion cells, when compound semiconductors such as InGaAs or InAlAs are grown, we are able to generate not only predictable variations of individual layer thicknesses but also of their compositions. Such controlled gradients provide a pathway to the systematic study and optimization of material properties, which is a capability unique to MBE technology.

The SLs discussed here were grown at a low temperature of approximately 250~°C, as measured using band-edge thermometry (BET)~\cite{johnson1993semiconductor}. At the center of the wafer, the doping concentration of Be was $4\times10^{18}~\text{cm}^{-3}$, and the growth rates were $2.5~\text{\AA} /\text{s}$ for the InGaAs and InAlAs layers.  Before the growth process, the substrate was first outgassed in the load lock chamber at 200~°C for 120 minutes, followed by heating to 300~°C in the preparation chamber for 60 minutes. Finally, native surface oxide was thermally removed by annealing at 530~°C for 10-20 minutes under arsenic ($\text{As}_4$) flux in the main MBE chamber. 

To systematically explore the properties of the SLs, three quarters of 3-inch Be-doped LT-In$_x$Ga$_{1-x}$As/In$_y$Al$_{1-y}$As SLs were grown. These SLs span a broad range of structural parameters, including strain, quantum well, and barrier compositions, as well as their respective thicknesses. The structural parameters of the grown SL samples are summarized as follows:

\begin{itemize}
    \item Sample A: The center compositions are $x_0 = 0.53$ and $y_0 = 0.52$. The composition ranges from $x = 0.49$ to $0.62$ and from $y = 0.45$ to $0.58$.
    
    \item Sample B: The center compositions are $x_0 = 0.65$ and $y_0 = 0.4$. The composition ranges from $x= 0.61$ to $0.72$ and from $y = 0.34$ to $0.46$.
    
    \item Sample C: The center compositions are $x_0 = 0.75$ and $y_0 = 0.32$. The composition ranges from $x= 0.72$ to $0.81$ and from $y = 0.27$ to $0.38$.
\end{itemize}

\begin{figure}
\centering
\includegraphics[width=\textwidth]{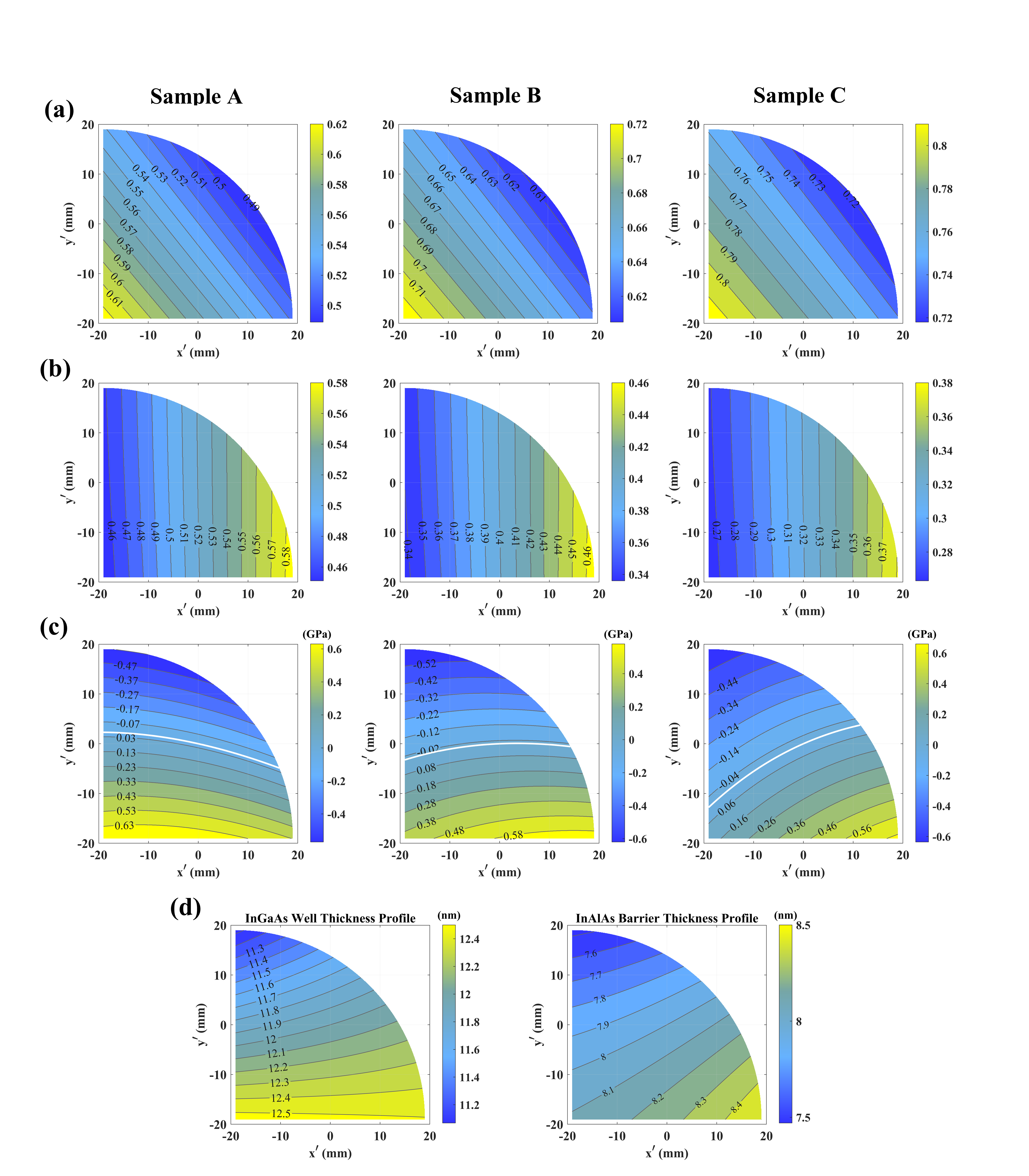}
\caption{(a) and (b) In composition profiles in the quantum wells and barriers, respectively, for Sample A, Sample B, and Sample C. (c) Average in-plane stress distribution ($\bar{\sigma_{\parallel}}$) across Sample A, B, and C. The bright curve line indicates the zero-stress reference. (d) Thickness profile of the quantum wells and barriers.}
\label{fig:3}
\end{figure}

Figures~\ref{fig:3}.a and~\ref{fig:3}.b display the distribution profiles of the indium mole fraction in the wells and barriers, respectively, while their corresponding thicknesses are shown in Figure~\ref{fig:3}.d. Furthermore, Figure~\ref{fig:3}.c shows the average in-plane stress distribution profile for grown samples. Notably, strain balancing is not restricted to a single design point but instead occurs along a trajectory where the average in-plane stress per period is maintained at zero. This stationary growth method enabled the fabrication of SLs with a diverse set of structural parameters on each substrate, providing a comprehensive experimental platform for investigating photoconductive properties across a broad range of parameters.

\section{Material Characterization}

\begin{figure}[t!]
\centering
\includegraphics[width=\textwidth]{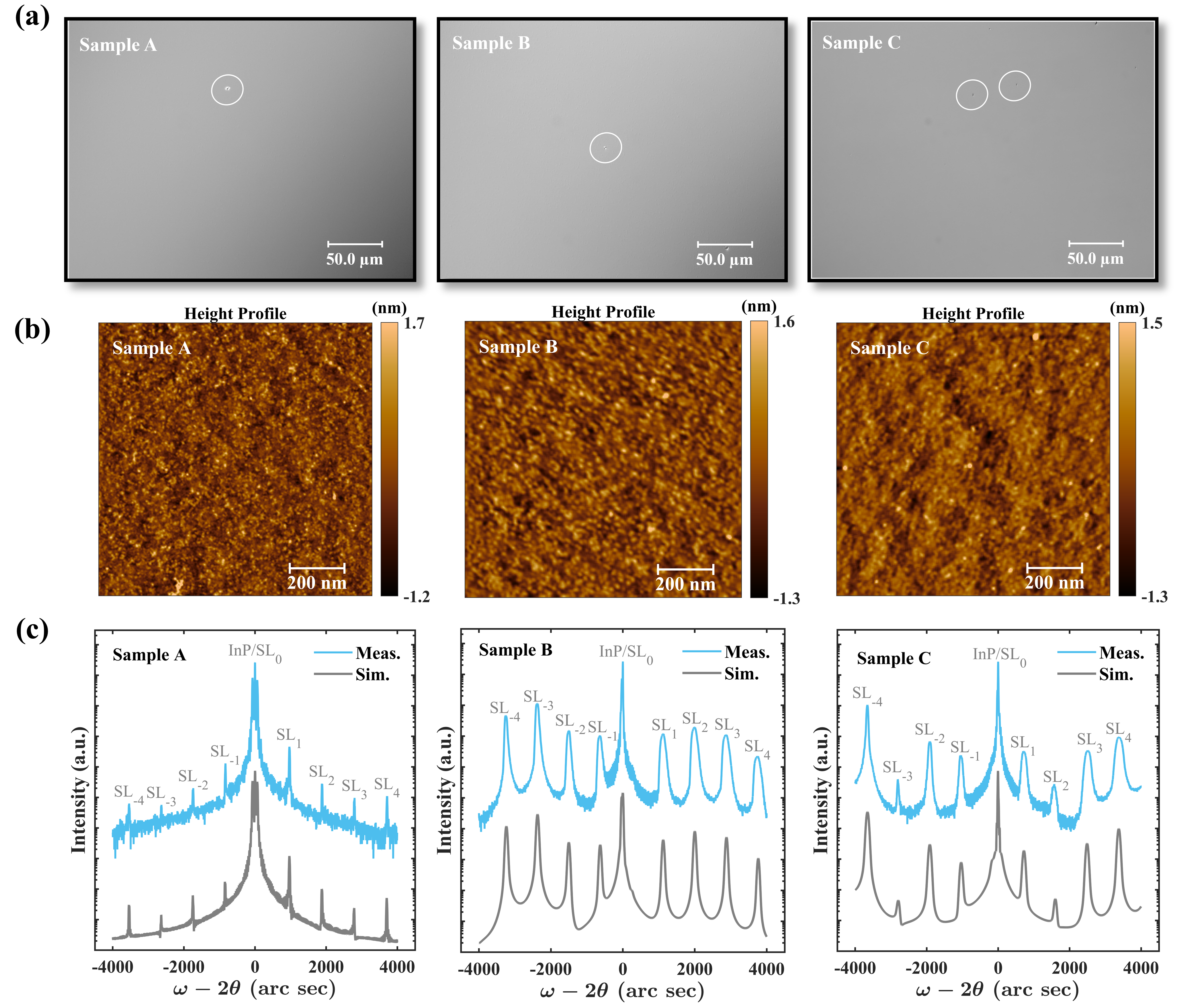}
\caption{(a) Nomarski DIC images of Sample A, B and C with scale markers. The highlighted circles indicate the location of point defects. (b) AFM images with a $1 \times 1 ~  \text{µm}^2$ scan range of height profile for Sample A (RMS roughness of 0.4~nm), Sample B (0.42~nm), and Sample C (0.36~nm). (c) X-ray diffraction scans on a logarithmic scale for the (400) reflection, along with corresponding RADS dynamical simulations for Samples A, B, and C. Each increment in the logarithmic scale represents one order of magnitude.} 
\label{fig:4}
\end{figure}

The surface morphology of the grown SLs was inspected using a Nikon Optiphot-66 Nomarski DIC microscope equipped with a SPOT digital camera. As shown in Figure~\ref{fig:4}.a, the surface exhibits a smooth morphology, with only a few isolated point defects observed. The scanned areas were deliberately chosen to include at least one defect, ensuring proper focus during inspection. Further surface morphology and roughness analysis were performed using a Bruker Dimension Icon AFM. In Figure~\ref{fig:4}.b  $1\times1~\text{µm}^2$ AFM scans for the three samples are presented. For Sample A, the height profile reveals a root mean square (RMS) surface roughness of 0.4~nm. Similarly, Sample B exhibits an RMS surface roughness of 0.42~nm, while Sample C shows an RMS value of 0.36~nm. These results confirm atomically smooth surfaces across all samples.

\begin{figure}[t!]
\centering
\includegraphics[width=\textwidth]{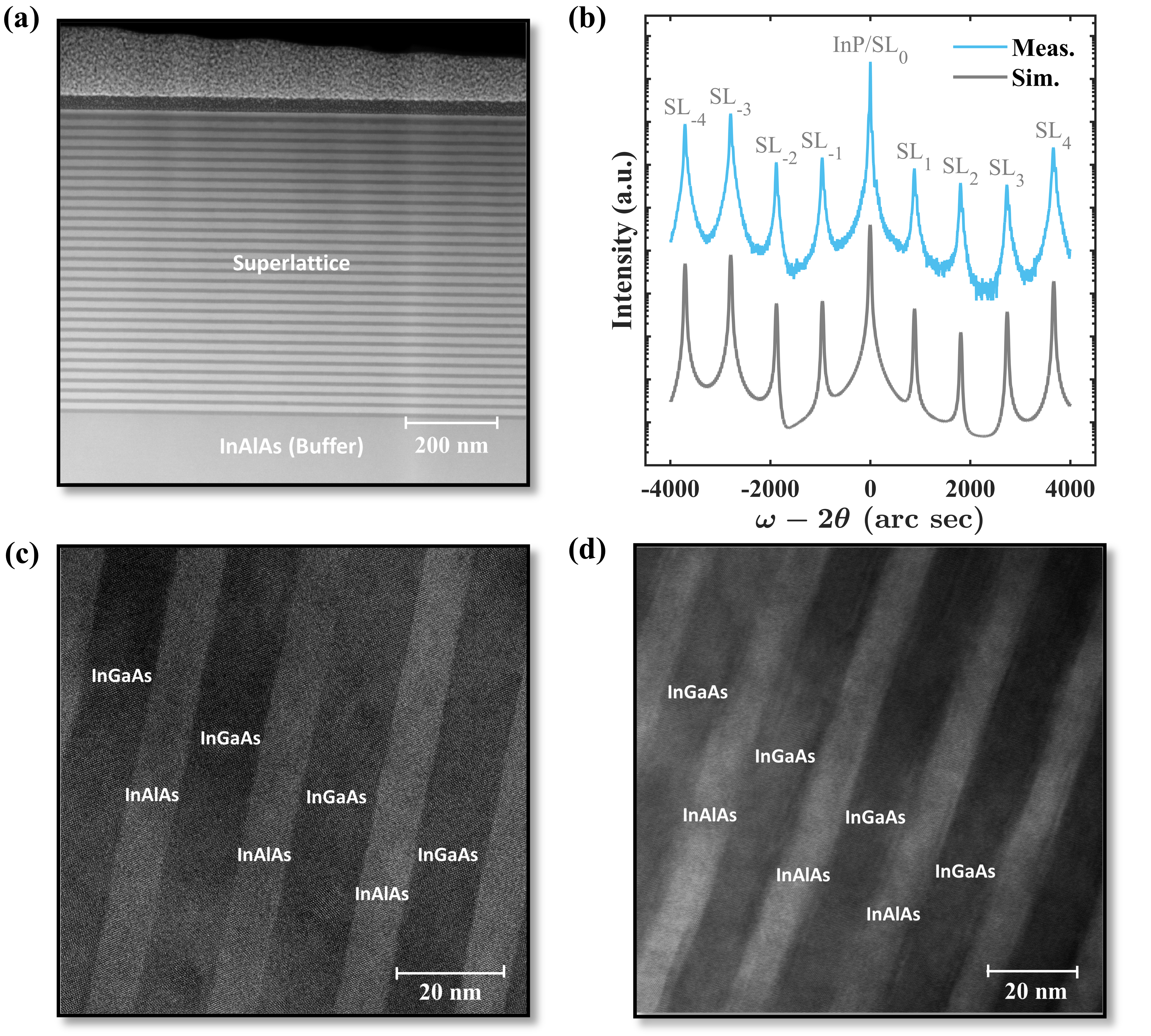}
\caption{(a) Dark-field STEM image of 30-period Be-doped InGaAs/InAlAs strain-balanced SL. (b) X-ray diffraction scans on a logarithmic scale for the (400) reflection, along with corresponding RADS dynamical simulations for an undoped strain-balanced sample with $x=$0.7 and $y=$0.3 (c) and (d) TEM image of the undoped and Be-doped InGaAs/InAlAs strain-balanced SL interfaces, respectively.}
\label{fig:5}
\end{figure}

HRXRD was used to assess the structural quality of the grown SLs. This technique provides insights into structural parameters such as layer thickness, composition, strain, and overall crystal quality by analyzing diffraction patterns. All HRXRD data were analyzed using Rocking-curve Analysis by Dynamical Simulation (RADS) software. The model indicates the presence of approximately one monolayer (5 $\text{\AA}$) InAs layer at the interface between the InAlAs buffer layer and the InP substrate, due to the displacement of phosphorous by arsenic on the InP surface during oxide desorption. Figure~\ref{fig:4}.c shows the $\mathrm{\omega}-2\mathrm{\theta}$ double-axis scans around InP (400) reflection, obtained using a JV-QC3 Bruker system. Several key observations arise from the HRXRD pattern of the superlattice structures. The most intense peak at zero arcseconds corresponds to the (400) Bragg reflection of the InP substrate. The angular position of the zero-order SLs peak ($\mathrm{SL}_0$) represents the average lattice constant of the InGaAs/InAlAs SLs stack, calculated as $\bar{\epsilon_{\perp}}= (\epsilon_w\,l_w+\epsilon_b\,l_b)/(l_w+l_b)$. In these samples, the zero-order SLs peak is partially masked by the peaks of the InP substrate and the InAlAs buffer layer. The presence of sharp periodic SL satellite peaks confirms the high structural quality of the grown SLs, indicating uniformity in layer thickness and composition. However, peak broadening is observed in Be-doped strain-balanced SLs, probably caused entirely by Be-induced interface interdiffusion. In comparison, similar undoped samples exhibit sharp satellite peaks, as shown in Figure~\ref{fig:5}. b, highlighting the influence of doping and growth conditions on structural quality. Additionally, significant incorporation of excess As into the crystal lattice during low-temperature growth contributes to lattice expansion~\cite{yano2001pump}, thereby shifting the SL satellite peaks.

To further evaluate the crystalline quality of the grown strain-balanced SLs, we employed S/TEM to characterize the epitaxial layers. The TEM lamelas of the samples were prepared using standard focused ion beam (FIB) techniques with a Zeiss Auriga 40 SEM / FIB instrument. Cross-sectional scanning TEM (STEM) analysis revealed high crystalline quality and smooth interfaces, as shown in the dark-field STEM micrograph in Figure~\ref{fig:5}.a.  TEM images of the undoped and Be-doped strained interfaces are presented in Figure~\ref{fig:5}.c and~\ref{fig:5}.d, respectively. 
These images confirm that incorporation of Be doping during low-temperature MBE growth of strain-balanced InGaAs/InAlAs SLs induces impurity-assisted interface disordering. To the best of our knowledge, this represents the first reported observation of Be-doping induced interdiffusion of the LT-InGaAs/InAlAs interfaces. Such intermixing at the interfaces is expected to significantly impact carrier transport and recombination dynamics. A similar interdiffusion phenomenon has been reported for Si ion implantation, where the quantum well properties were altered and a blue shift in the photoluminescence (PL) spectra was induced~\cite{yamamura1994defect}.
Despite the observed interdiffusion, no misfit dislocations, stacking faults, or twins were detected, and no evidence of phase separation in InGaAs or InAlAs was found throughout the examined lamella. These findings confirm the high structural integrity of the SLs, highlighting both the precision of the growth process and the robustness of the strain-balanced design.

\section{Photoconductive Properties}

This section investigates the photoconductive properties of the grown SLs. When an optical pump signal with a photon energy exceeding the semiconductor band-gap energy illuminates the material, electron-hole pairs are generated through optical absorption. In this process, electrons are excited into the conduction band, leaving behind holes in the valence band. These photoexcited carriers then diffuse and drift through the photoconductor until they are eventually trapped and recombine. As the photoexcited carriers move within the photoconductor, they undergo various relaxation processes before returning to the thermodynamic equilibrium, which influence the overall carrier dynamics. These relaxation mechanisms include direct radiative interband recombination as well as nonradiative trap-assisted and Auger recombination. Because the radiative interband recombination in the grown SLs occurs on a nanosecond time scale, ultrafast nonradiative recombination mechanisms dominate in low-temperature-grown Be-doped InGaAs/InAlAs SLs. These nonradiative pathways, primarily driven by defect-assisted trapping and recombination, lead to rapid decay of the photoexcited carrier population. Together, these recombination mechanisms produce multiexponential carrier dynamics, with each decay component characterized by a distinct photoexcited carrier lifetime, $\tau$. The temporal behavior of carrier densities in doped superlattices can be described by a rate equation~\cite{Int20,vignaud2002electron}, which accounts for carrier generation, trapping, and recombination. 
By accounting for doping, defect states, and injection-dependent recombination rates, the rate equation provides a useful framework for analyzing photoexcited carrier dynamics. Understanding these recombination mechanisms is essential for further optimization of the SL design towards improved performance of THz photoconductive antennas.

\subsection{Photoexcited Carrier Dynamics}

\begin{figure}
\centering
\includegraphics[width=\textwidth]{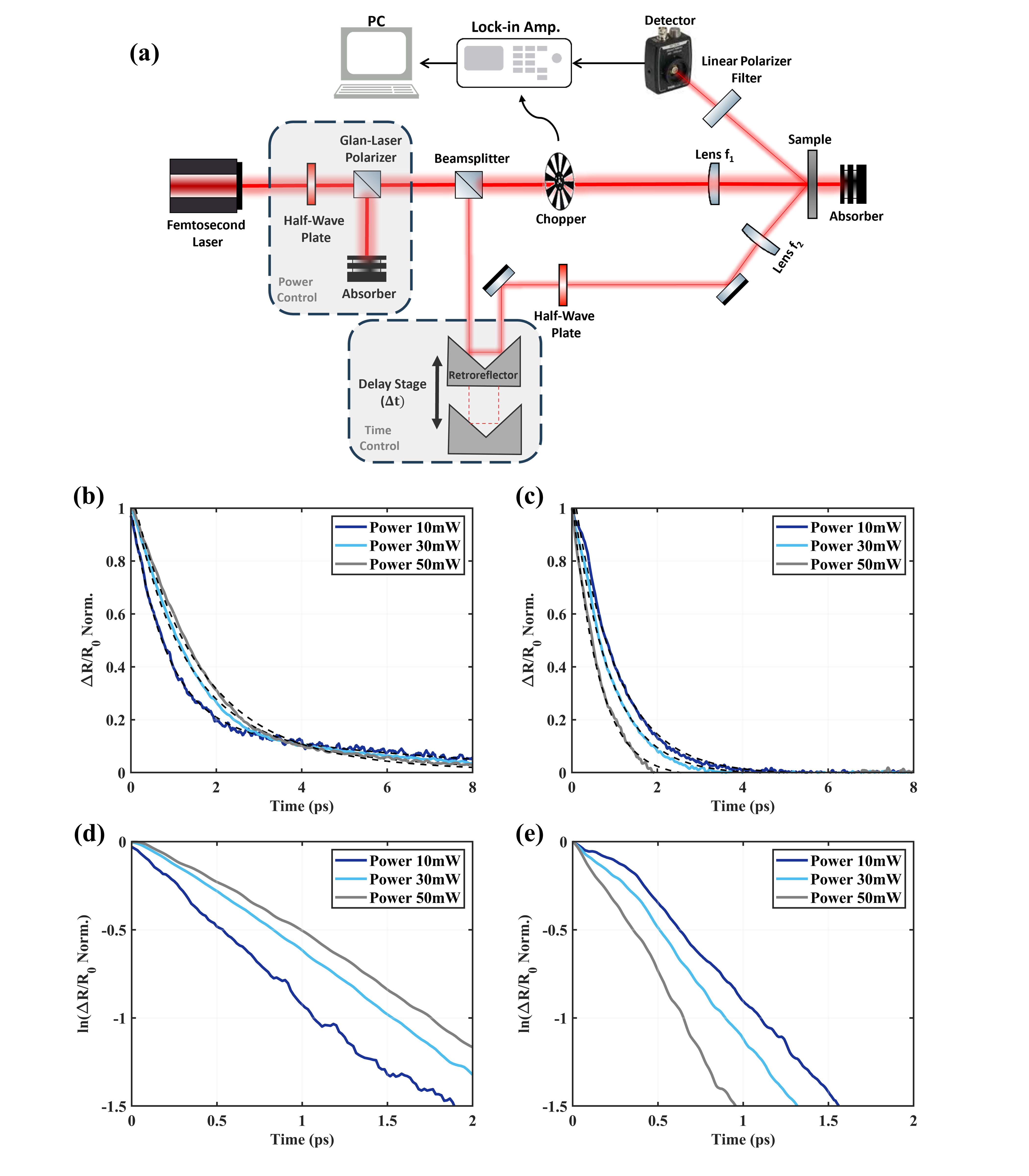}
\caption{(a) Schematic diagram of the time-resolved differential reflectivity (DR) pump-probe spectroscopy setup. (b) and (c) Normalized differential reflectivity signals for increasing pump pulse powers of 10 mW, 30 mW and 50 mW at two different measurement points of strained-balanced samples at $x=0.7$ \& $y=0.34$ and at $x=0.8$ \& $y=0.26$, respectively. (d) and (e) show the corresponding natural logarithm of differential reflectivity signals (b) and (c), respectively.}
\label{fig:6}
\end{figure}

To investigate subpicosecond photoexcited carrier dynamics in the grown SLs, an optical pump-probe transient reflectivity method was employed using ultrafast 1550~nm femtosecond laser excitation. Pump-probe spectroscopy is a widely used optical technique that circumvents electronic limitations, thus enabling insight into picosecond-scale relaxation dynamics of photoexcited carriers in semiconductor structures. In this setup, the optical pump pulses disturb the equilibrium state, while time-delayed probe pulses measure the evolution of the sample's transient reflectivity. By probing photoexcited changes at different time delays and averaging multiple measurements, the time-resolved differential reflectivity (DR) pump-probe signal is reconstructed. Figure~\ref{fig:6}.a illustrates the schematic of the time-resolved pump-probe spectroscopy setup in a reflection geometry. An erbium-doped ultrafast pulsed femtosecond laser, centered at 1550~nm wavelength, generated 100 fs optical pulses at an 80 MHz repetition rate with an average power up to 250 mW, which was used to study the transient reflectivity.

The laser beam is split into pump and probe components using a pellicle beamsplitter. The pump beam, which contains most of the optical power (with a typical 10:1 ratio relative to the probe beam), passes through a chopper operating at a frequency of $f_{ch}= 430$ Hz and then is focused on the sample surface by a lens with a focal length of $\text{f}_{1}= 7.5$ cm. This pump beam generates a non-equilibrium carrier population. The probe beam, delayed by a motorized delay line, is focused by a lens with a focal length of $\text{f}_{2}= 5$ cm to coincide with the focused pump beam on the superlattices. To ensure that the pump spot is larger than the probe spot, the pump lens has a longer focal length than the probe lens, following the principles of beam profile optimization~\cite{prasankumar2016optical}. The reflected probe beam, incident at an angle of 60°, is collected by a DET10C2 biased InGaAs photodetector, which converts the optical signal into an electrical signal. To eliminate interferences, a half-wave plate rotates the probe beam's polarization by 90°, cross-polarizing it with the pump beam, while a linear polarizer filter isolates it from any unwanted pump reflections. By scanning the delay stage, the electrical signal is recorded at each step using an SR850 digital lock-in amplifier, enabling reconstruction of the optical pump-probe transient reflectivity.

The measured optical pump-probe transient reflectivity signals, $\mathrm{\Delta R/R_0}$ (where $\mathrm{\Delta R(t)}$ is the change in probe reflectivity and $\mathrm{R_0}$ is the initial signal before pump excitation), typically show a subpicosecond increase in the signal, followed by a fast exponential initial decay, followed by a slower decay. These dynamics provide valuable insights into the multiple carrier trapping and recombination processes occurring within the material. The transient reflectivity responses were fitted using the following bi-exponential decay equation:
\begin{equation}
\frac{\Delta R(t)}{R_0} = A  e^{-t/\tau_1} + B e^{-t/\tau_2}+C
\label{Eq.5}
\end{equation}
Here, $\mathrm{\tau_1}$ and $\mathrm{\tau_2}$ represent carrier lifetimes corresponding to trapping/recombination processes. The coefficients $\mathrm{A}$, $\mathrm{B}$, and $\mathrm{C} $ are the fitting constants, where $\mathrm{C} $ may account for the residual photoexcited carrier populations remaining after recombination. We expect the dominant decay mechanisms to be primarily affected by the photoexcited carrier density and defect concentration. 

At low injection levels (the "unsaturated trap" regime), where the density of available trap sites exceeds the number of photoexcited carriers, carrier lifetimes are primarily governed by electron capture from the conduction band by positively ionized arsenic antisite defects ($\mathrm{As}^{+}_{\mathrm{Ga}}$), followed by electron-hole recombination~\cite{Int20}.  
Be doping (p-type) compensates and ionizes these defects, strongly influencing carrier concentration and dynamics. Hence, in the LT-InGaAs layers, the electron trapping time is largely determined by the concentration of positively ionized arsenic antisite defects ($\mathrm{As}^{+}_{\mathrm{Ga}}$), which depends directly on beryllium doping levels~\cite{Int20,Int21}. The effect of pump power on the photoexcited carrier lifetimes in Sample B, measured at $x=$0.7 and $y=$0.34, is shown in Figure~\ref{fig:6}.b. The results indicate a trend of increasing carrier lifetimes with higher pump power, consistent with electron trapping remaining the dominant recombination mechanism~\cite{Int20}.

Moreover, Shockley-Read-Hall (SRH) recombination, which proceeds via defect states within the bandgap, is also important in low-temperature-grown materials. Excess arsenic incorporation and the formation of neutral arsenic antisite defects introduce deep-level electron-hole recombination centers. These defects facilitate SRH recombination and significantly reduce the recombination time $\mathrm{\tau_2}$.  The high density of these defect states, particularly in Be-doped low-temperature-grown InGaAs/InAlAs SLs, enhances both carrier trapping and recombination rates. 

At high injection levels (the "saturated trap" regime), where the density of photoexcited carriers exceeds the available trap states, these trap states become progressively filled, reducing their ability to capture additional photoexcited carriers. As a result, the recombination dynamics shift toward high carrier density-dependent processes.  The observed initial faster decay with increasing pump power is hypothesized to be dominated by nonradiative processes, primarily carrier-carrier scattering, hot-carrier effects and trap-assisted Auger recombination, all strongly dependent on carrier density, defect state concentrations, and energy redistribution dynamics~\cite{shah2013ultrafast,vignaud2002electron,joschko2000ultrafast,othonos1998probing}. The effect of pump power on the photoexcited carrier lifetimes in Sample C, measured at  x=0.8 and y=0.26 point, is shown in Figure~\ref{fig:6}.c. These results reveal a trend of decreasing carrier lifetimes with increasing pump power, indicating a transition to high-density-dependent recombination mechanisms. This behavior strongly supports the hypothesis that carrier-carrier scattering and hot-carrier effects dominate the initial relaxation dynamics through momentum randomization and energy redistribution among carriers, facilitating ultra-fast relaxation and accelerating recombination. Furthermore, trap-assisted Auger recombination can further contribute to this regime. ~\cite{vignaud2002electron,shah2013ultrafast,joschko2000ultrafast,othonos1998probing}. 

An additional scenario for the observed rapid initial decay could involve the formation of transient excitons via direct hot-exciton photogeneration or electron collisions through Auger processes, with subsequent exciton-exciton and exciton-phonon scattering facilitating sub-picosecond relaxation~\cite{nie2020transient}. Nevertheless, considering the rapid dissociation of excitons in defect-rich, low-temperature-grown Be-doped superlattices, purely excitonic mechanisms generally do not exhibit the power-dependent accelerated relaxation observed in our results, making it unlikely that exciton interactions alone account for the fast initial decay.

The results highlight the complexity of carrier dynamics in Be-doped, low-temperature-grown SLs, where defect-assisted trapping and recombination combine with high-density carrier interactions to shape the observed behavior. However, the precise interplay of these mechanisms remains unclear without advanced modeling. Further experimental studies are also required to fully understand the dynamics under high-injection conditions, especially in the presence of substantial doping-induced defect states. 

\begin{figure}
\centering
\includegraphics[width=0.93\textwidth]{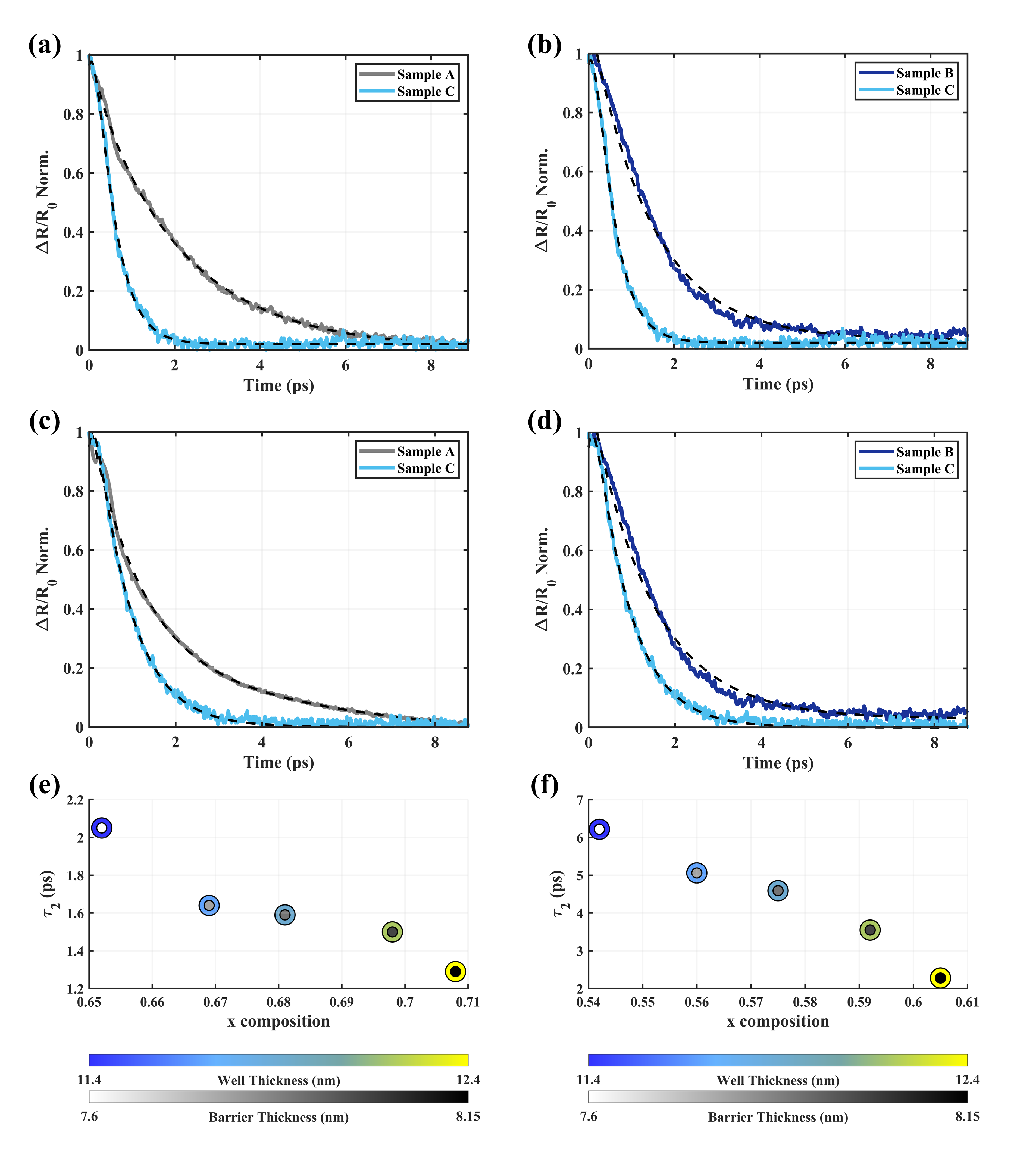}
\caption{Normalized differential reflectivity signals comparing the lattice-matched SL (sample 1 at $x=0.53 $ \& $y=0.52$) and strained-balanced SLs (sample 2 at $x=0.7$ \& $y=0.34$, sample 3 at $x=0.8$ \& $y=0.26$). (a) and (b) correspond to a pump power of 30 mW, while (c) and (d) correspond to 15 mW for all samples. The signal decay times decrease for higher well composition (In$_x$) and lower barrier composition (In$_y$). Dashed black lines are exponential fits to the signals. (e) and (f) show carrier lifetimes ($\tau_2$) as a function of well composition (In$_x$), where the color scales indicate well and barrier thicknesses. Data points correspond to measurements at $y=0.35$ and $y=0.47$, respectively.}
\label{fig:7}
\end{figure}

This study focuses on a systematic investigation of photoexcited carrier lifetimes across a broad range of structural parameters in the same Be-doped, low-temperature-grown SLs. All measurements were performed at room temperature, with the time-zero point defined as the moment when the pump and probe pulses simultaneously reached the SL surface. Normalized pump-probe DR signals, measured at various points and input power levels throughout the three samples, each with distinct well and barrier structural parameters, are presented in Figure~\ref{fig:7}. The dashed black lines indicate exponential fits to the DR signals, providing a visual representation of the carrier dynamics. The detailed comparison of the photoexcited carrier lifetime at different indium mole fractions in the wells ($x$-composition) and barriers ($y$-composition) under two pump power levels is summarized in Table \ref{tbl:lifetime_comparison}. The measurement points were carefully selected to ensure that the well and barrier thickness were nearly identical, thus isolating the effects of compositions on the carrier lifetime. The extracted carrier lifetimes demonstrate a clear dependence on structural parameters, particularly the indium compositions in InGaAs wells. Figures~\ref{fig:7}.e and~\ref{fig:7}.f show the fitted carrier lifetimes ($\tau_2$) as a function of the In composition in the wells ($x$).
Shorter carrier lifetimes are observed when the x-composition increases (larger compressive strain in the wells), whereas the y-composition remains the same. This finding illustrates the influence of strain on the ultrafast carrier trapping/recombination dynamics. The shortest observed photoexcited carrier lifetime ($\tau_1=0.2 \pm  0.02$ ps and $\tau_2=0.58 \pm  0.02$ ps) at 30 mW pump power were recorded in the strain-balanced SLs with the highest compressive strain in the wells ($x=0.8$) and the highest tensile strain in the barriers ($y=0.26$).

\begin{table} [t!]
  \caption{Comparison of photoexcited carrier lifetimes at different In compositions in wells (\( x \)) and barriers (\( y \)) under pump power 15 mW and 30 mW. The measurement points are selected to have almost same wells and barriers thicknesses ($\mathrm{l_w=12.1 \, nm,l_b=8 \, nm}$). }
  \label{tbl:lifetime_comparison}
  \setlength{\tabcolsep}{8pt}
  \renewcommand{\arraystretch}{1.2}
  \resizebox{\textwidth}{!}{
  \begin{tabular}{lccccccc}
    \hline
     \multicolumn{2}{c}{Composition} & \multicolumn{3}{c}{Power= 15 mW} & \multicolumn{3}{c}{Power= 30 mW} \\
        \cline{3-8}
     \( x \) & \( y \) & $\tau_1$ (ps) & $\tau_2$ (ps) & RMS Error & $\tau_1$ (ps) & $\tau_2$ (ps) & RMS Error \\
    \hline
     0.53 & 0.52 & 1.19 $\pm$ 0.02 & 8.33 $\pm$ 0.02 & 0.0004 & 1.55 $\pm$ 0.1 & 2.96 $\pm$ 0.3 & 0.0003 \\
     0.6  & 0.46 & 0.64 $\pm$ 0.02 & 5.82 $\pm$ 0.02 & 0.0006 & 0.77 $\pm$ 0.03 & 2.28 $\pm$ 0.1 & 0.0004 \\
     0.65 & 0.4  & - & 1.83 $\pm$ 0.003 & 0.0003 & - & 2.02 $\pm$ 0.002 & 0.0004 \\
     0.7  & 0.35 & - & 1.402 $\pm$ 0.02 & 0.0007 & - & 1.504 $\pm$ 0.02 & 0.0007 \\
     0.8  & 0.26 & 0.2 $\pm$ 0.05 & 0.8 $\pm$ 0.05 & 0.0004 & 0.2 $\pm$ 0.02 & 0.58 $\pm$ 0.02 & 0.0004 \\
    \hline
  \end{tabular}}
\end{table}

\subsection{Absorption Coefficient}

\begin{figure}[t!]
\centering
\includegraphics[width=0.95\textwidth]{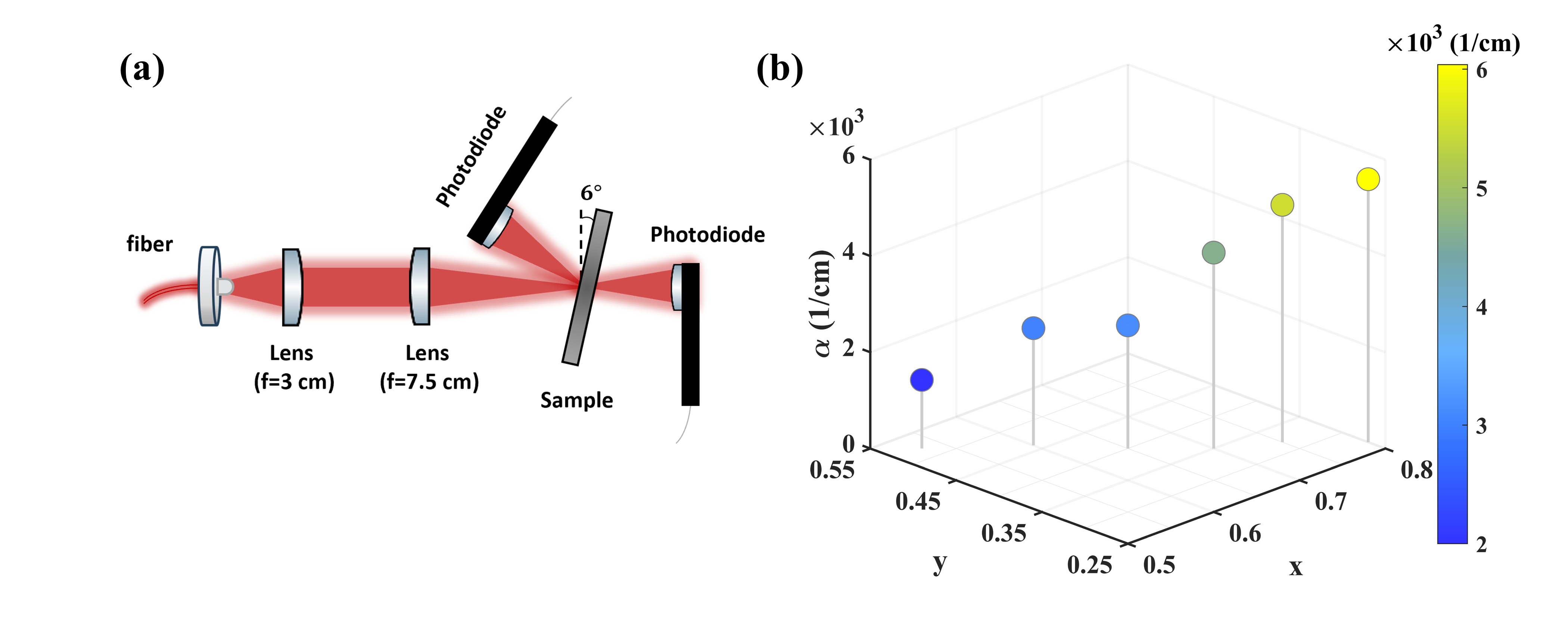}
\caption{(a) Schematic representation of the absorption spectroscopy setup. (b) Averaged optical absorption coefficient ($\alpha$) over the wavelength range $\lambda = 1550 \pm 10$~nm for various indium mole fractions in wells ($x$-composition) and corresponding strain-balanced indium mole fractions in barriers ($y$-composition). }
\label{fig:8}
\end{figure}

Optical absorption spectroscopy with continuous wave (CW) laser excitation was performed to determine the absorption coefficient $\alpha$ (1/cm). A commercial Fabry–Pérot laser diode, emitting at $\lambda_c = 1550 \pm 10$~nm, was selected as the excitation source. The sample was positioned at a 6º incidence angle relative to the laser beam. Photodiode detectors were placed to measure the reflected and transmitted power at 12º and 180º relative to the incident beam, respectively. As shown in the schematic representation in Figure~\ref{fig:8}, the average reflected and transmitted powers at this wavelength were recorded using a high-resolution photodiode (Model S132C), which offers a sensitivity of 1 nW, ensuring precise power measurements. 

Because the coherence length of the laser light significantly exceeds the thickness of the superlattice yet remains much smaller than that of the substrate, all laser waves within the superlattice interfere coherently, whereas interference in the substrate is assumed to be incoherent. Furthermore, the SI-InP:Fe substrate is assumed to be transparent at 1550~nm, which simplifies the computational analysis. Under these conditions, the total power reflection and transmission were calculated, with detailed derivations provided in the \textbf{Supporting Information}. 

The calculated absorption coefficient $\alpha$ is shown in Figure~\ref{fig:7}.b for different indium mole fractions in the wells ($x$-composition) and the corresponding strain-balanced indium mole fractions in the barriers ($y$-composition). Our sensitivity analysis indicates that minor variations in the thickness of the superlattice do not significantly affect the calculated absorption coefficients. While treating the InP substrate and InAlAs buffer layers as effectively lossless and using an average refractive index for the superlattice may oversimplify its actual index profile, the resulting uncertainty remains small.

The results indicate that increasing the $x$-composition in strain-balanced SLs, which feature compressively strained InGaAs layers and tensile strained InAlAs layers, leads to higher optical absorption compared to lattice-matched SLs. A higher indium mole fraction in the InGaAs layers reduces the bandgap energy, thereby boosting optical transitions and enabling more efficient photon absorption. This bandgap reduction effectively increases the absorption coefficient, an essential factor for maximizing photon-to-carrier conversion efficiency in photoconductive devices.

\subsection{Carrier Mobility}

In addition to characterizing the absorption coefficient and carrier lifetime, the transport properties of the grown samples were evaluated using Hall effect measurements. Room-temperature Hall effect measurements were performed in van der Pauw geometry to determine Hall mobility, sheet resistance, and carrier concentration of the samples.  
The wafers were cleaved into $7 \text{mm}\times5\text{mm}$ pieces,  minimizing the variation in structural parameters within each piece. Indium contacts were fabricated at each corner, and effective transport parameters were measured. 
The extracted values were correlated with the center compositions of each sample, as summarized in Table \ref{tbl:hall_properties}. None of the samples studied received post-growth annealing. 

All pieces exhibited n-type conductivity, with a sheet carrier concentration ranging from $10^{11}$ to $10^{13} ~ \text{cm}^{-2} $ . The results indicate that strain-balanced SLs with compressively strained InGaAs layers and tensile strained InAlAs layers exhibit higher mobility compared to their lattice-matched counterparts. The highest measured mobility in grown strain-balanced SLs is  $ \mu=682.6 ~ \text{cm}^2\text{/Vs}$, measured at $x=0.65$ and $y=0.38$ point. In comparison, the corresponding lattice-matched SLs exhibited a lower mobility of $ \mu= 408.5 ~ \text{cm}^2\text{/Vs}$. These values compare favorably with the values reported in the literature for lattice-matched SLs~\cite{Int22}. 

\begin{table} [t!]
  \caption{Hall Effect Properties of Grown Superlattice Samples }
  \label{tbl:hall_properties}
  \setlength{\tabcolsep}{8pt} 
  \renewcommand{\arraystretch}{1.2} 
  \resizebox{\textwidth}{!}{
  \begin{tabular}{lcccccc}
    \hline
    Sample & $x$& $y$  &$\mu_e$ (cm$^2$/Vs)& $n_e$ (cm$^{-2}$) /well& $\text{R}_s\,$(k$\Omega/\Box$)&Carrier Type\\
    \hline
    Reference~\cite{Int22}& 0.53 & 0.52  &517& 8.1 $\times$ 10$^{9}$& 15&n-type\\
 Sample A& 0.53 & 0.52  & 408.5& 2.48$\times$ 10$^{10}$& 20.7&n-type\\
 & 0.61& 0.45& 549& 3.05$\times$ 10$^{10}$& 12.4&n-type\\
    Sample B& 0.65& 0.38&682.6& 3.83 $\times$ 10$^{10}$& 8.09&n-type\\
 & 0.71& 0.33& 623.6& 1.14 $\times$ 10$^{11}$& 3.02&n-type\\
 Sample C& 0.75& 0.28& 601.3& 1.49 $\times$ 10$^{11}$& 2.13&n-type\\
 & 0.8& 0.25& 532.3& 3.66$\times$ 10$^{11}$& 1.06&n-type\\
     \hline
  \end{tabular}}
\end{table}

\section{Discussion}

We systematically investigated the interplay of structural parameters in Be-doped low-temperature-grown InGaAs/InAlAs SLs by employing strain-balancing technique under stationary-growth condition in MBE. Although strain balancing itself is an established method for enabling strained, yet pseudomorphic epitaxial growth, our stationary growth approach uniquely introduces lateral gradients in layer composition, strain, and thickness. This method provides an extensive experimental platform on a single wafer for the precise tuning of photoconductive properties across a wide parameter space. Exploring these parameters provides deeper insight into how structural variations influence electronic band structure, optical absorption, carrier transport, and dynamics. The structural uniformity confirmed through S/TEM and HRXRD data demonstrates that, despite intentionally introducing substantial strain, the superlattices consistently maintain high crystalline quality.

Strain balancing, achieved through pairing compressively strained InGaAs wells with tensile strained InAlAs barriers, facilitates high-quality epitaxial growth and significantly modifies the electronic band structure and optical absorption properties. Increasing the In content in the bulk InGaAs increases its lattice constant and results in a decrease in the forbidden gap. This, in turn,  decreases the effective mass of electrons in the conduction band, increasing their mobility. However, it is not immediately obvious what would happen with the conduction and valence bands in InGaAs/InAlAs SLs, where InGaAs is under compressive and InAlAs is under tensile stain.  Indeed, even though the relaxed lattice constant of InGaAs would increase, the unit cell lateral size remains the same in a strain-balanced SL and the transport of electrons and holes will take place in the lateral dimension.  We show with the 8-band k.p modeling that in this situation the superlattice bandgap is nevertheless reduced and the electron mass slightly decreased. Moreover, the hybridization of heavy and light holes due to biaxial strains in the layers leads to a strong increase in the curvature of the top valence band at $k_{\parallel}$ = 0, thus a considerable decrease in the effective mass of the holes. It is clear that compressively strained wells are very beneficial for the lateral transport of photoexcited carriers in the superlattice, in part due to increased mobility of both electrons and holes. In addition, the conduction and valence band discontinuities in the strain-balanced SLs discussed here are both almost twice larger than that in the lattice-matched SL. This increases carrier confinement and reduces scattering in InAlAs barriers, which are expected to be more defective than InGaAs in the wells. Moreover, a smaller band gap in the strain-balanced SL is expected to enhance photon absorption and improve photon-to-carrier conversion efficiency. All these effects lead to enhancement of parameters that are of key importance for efficient THz PCAs. 

Although these mechanisms collectively enhance mobility,  one can argue that an increase in scattering on "harder" interfaces may degrade the carrier transport, depending upon specific strain and doping conditions. Yet, the Be-induced intermixing reported here may counteract this by effective "softening" of the interfaces, reducing their scattering efficiency. Clearly, an experimental verification of the above predictions is necessary. 

Indeed, in the studies reported here, Hall measurements confirm that strain-balanced SLs with compressively strained wells lead to a considerable increase in electron mobility compared to that of lattice-matched reference structures with similar doping levels. These mobility enhancements, coupled with experimentally confirmed increased absorption, contribute to improved THz photoconductive performance.

Time-resolved pump–probe spectroscopy demonstrates subpicosecond photoexcited carrier lifetimes in these low-temperature strain-balanced SLs doped with Be, with the fastest decay ($\tau_1\simeq0.2 $~ps) observed in structures with the greatest compressive strain in the wells and the greatest tensile strain in the barriers. These ultrafast lifetimes are consistent with enhanced defect-assisted recombination mechanisms, presumably facilitated by the high defect densities characteristic of low-temperature growth and intentional Be doping. However, the precise interplay among strains, arsenic antisite defects, and Be-induced intermixing at the interfaces remains to be thoroughly clarified. While strain itself may not directly dictate these recombination pathways, our findings indicate that strain balancing effectively tailors ultra-fast carrier dynamics, making these superlattices promising candidates for broadband and efficient pulsed THz detection. 

Given the observed ultrafast lifetimes, it is also important to further consider the impact of Be doping on the interface properties. S/TEM analysis reveals that Be induces intermixing at the strained InGaAs/InAlAs interfaces, which may influence carrier transport and recombination dynamics. Although further quantitative studies are needed to determine the exact implications of this interdiffusion on carrier transport and recombination, this finding highlights the complexity introduced by doping and emphasizes the need for careful consideration of interfacial properties in future strained-balance SL designs, particularly in the context of optimizing these superlattices for THz photoconductive applications.

\section{Conclusion}

In summary, we demonstrated that strain-balanced Be-doped InGaAs/InAlAs superlattices grown via stationary MBE provide a robust experimental platform for studying and systematically tuning photoconductive properties. The stationary growth method introduced controlled lateral gradients in composition, strain, and thickness across a single wafer, allowing a comprehensive exploration of structural parameters. Extensive characterization confirmed high crystalline quality, smooth surface morphology, and structural integbrity over a wide parameter range. Enhanced optical absorption at 1550~nm, coupled with improved carrier mobility under strain-balanced conditions, was experimentally confirmed and supported by theoretical modeling. Ultrafast carrier dynamics measurements demonstrated subpicosecond lifetimes, underscoring the potential of strain balancing combined with Be doping in tailoring ultrafast recombination dynamics in low-temperature-grown superlattices. Furthermore, direct structural evidence from transmission electron microscopy revealed significant Be-induced interface intermixing, an important new observation that merits further detailed exploration. These results collectively establish low-temperature MBE grown strain-balanced superlattices as promising candidates for optimized broadband and efficient THz photoconductive detectors operating at telecom-compatible wavelengths.

 \begin{acknowledgement}
This work was financially supported by the Natural
Sciences and Engineering Research Council of Canada (NSERC). The authors acknowledge insightful valuable discussions with Dr. Arash Rohani and Dr. Daryoosh Saeedkia. The authors also thank Dr. Sandra Gibson and Nicki Shaw at the University of Waterloo’s Quantum-Nano Fabrication and Characterization Facility (QNFCF) for their support with AFM and TEM characterization.



 \end{acknowledgement}

\begin{suppinfo}

The Supporting Information includes additional details on the optical absorption analysis of multilayered thin films on a thick substrate, the theoretical framework used for extracting absorption coefficient of InGaAs/InAlAs superlattices with transfer matrix method.


\end{suppinfo}

\bibliography{achemso-demo}

\end{document}